\documentclass[a4paper,11pt]{article}
\usepackage[normalem]{ulem}
\usepackage{pos}

\usepackage{floatrow}
\newfloatcommand{capbtabbox}{table}[][\FBwidth]

\usepackage{blindtext}

\usepackage[T1]{fontenc}
\usepackage[utf8]{inputenc}
\usepackage{babel}
\usepackage[font=small,labelfont=bf]{caption}

\newcommand{\pTf}{p_{\scriptscriptstyle T}}
\newcommand{\pTi}{p_{\scriptscriptstyle T}^{\rm \tiny initial}}

\newcommand{\be}{\begin{equation}}
\newcommand{\ee}{\end{equation}}

\title{Classification of quark and gluon jets in hot QCD medium with deep learning}

\author*[a,b]{Yi-Lun Du}
\author[c]{Daniel Pablos}
\author[a]{Konrad Tywoniuk}

\affiliation[a]{Department of Physics and Technology, University of Bergen,\\ Postboks 7803, 5020 Bergen, Norway}

\affiliation[b]{Department of Physics, University of Oslo,\\Sem Sælands vei 24, 0371 Oslo, Norway}

\affiliation[c]{INFN, Sezione di Torino, via Pietro Giuria 1, \\ I-10125 Torino, Italy}

\emailAdd{yilun.du@uib.no}

\abstract{Deep learning techniques have shown the capability to identify the degree of energy loss of high-energy jets traversing hot QCD medium on a jet-by-jet basis. The average amount of quenching of quark and gluon jets in hot QCD medium actually have different characteristics, such as their dependence on the in-medium traversed length and the early-developed jet substructures in the evolution. These observations motivate us to consider these two types of jets separately and classify them from jet images with deep learning techniques. We find that the classification performance gradually decreases with increasing degree of jet modification. In addition, we discuss the predictive power of different jet observables, such as the jet shape, jet fragmentation function, jet substructures as well as their combinations, in order to address the interpretability of the classification task.}

\FullConference{%
  *** Particles and Nuclei International Conference - PANIC2021 ***\\
  *** 5 - 10 September, 2021 ***\\
  *** Online ***
}

\begin{document}
\maketitle

\section{Introduction}
The deconfined QCD matter created in heavy-ion collisions has been seen to behave as a strongly-coupled viscous fluid 
commonly known as the quark-gluon plasma (QGP). 
Jets, collimated sprays of hadrons generated in a hard QCD process, are witnesses to the evolution of the QGP. Partons from the jet interact with the hot medium, losing energy and experiencing a variety of modifications, known as jet quenching phenomena. The observed high-energy hadrons and jets can therefore be used as probes with which to extract the transport properties of the QGP. 
In our previous work~\cite{Du:2020pmp,du2021jet}, we use deep learning techniques to estimate, on a jet-by-jet basis, the amount of energy loss, quantified through the variable $\chi \equiv \pTf/\pTi$ suffered by jets in the QGP. Here, $\pTf$ is the transverse momentum of a given jet in the presence of a medium with cone size $R$, and $\pTi$ is the transverse momentum of the \emph{same} jet had there been no medium, see \cite{Du:2020pmp} for further details. 
In this context, it would also be interesting to consider the species of the jet initiator. Quark- and gluon-initiated jets experience different evolutions, both in vacuum and in the medium. 
Their different microscopic processes lead to different splitting functions. The splitting angle as well as the phase space for the medium-induced radiative energy loss for gluon jets are larger on average. On the other hand, the in-medium traversed length dependence of quark and gluon parton energy loss is also different due to their different Casimirs. 
This motivates us to explore the differences between quark and gluon jets, and classify them. 
In the following, we first present their different quenching behavior. Then we use deep learning techniques to classify them and analyze the predictive power of the different jet observables. Finally, we summarize and conclude.

\section{Different quenching behavior}
\begin{figure}[b!]
\centering
\includegraphics[width=0.46\textwidth]{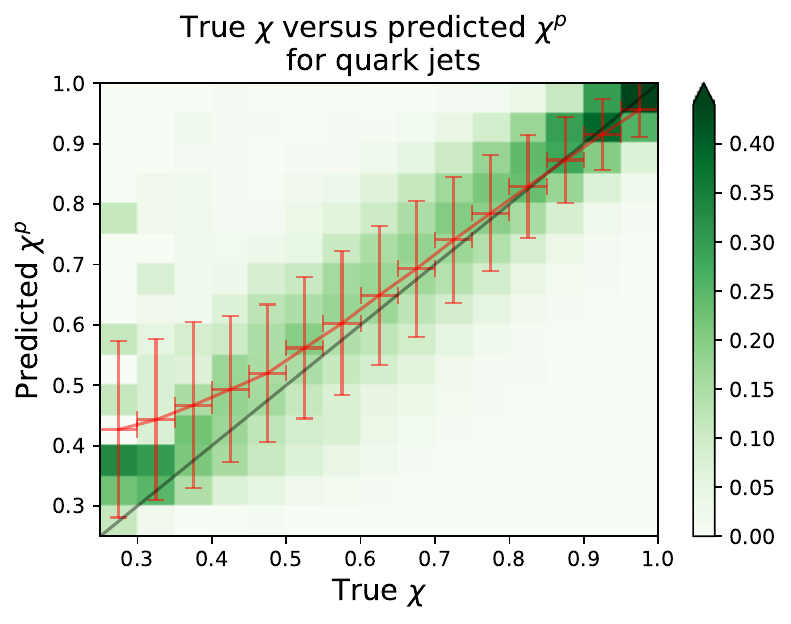}
\includegraphics[width=0.46\textwidth]{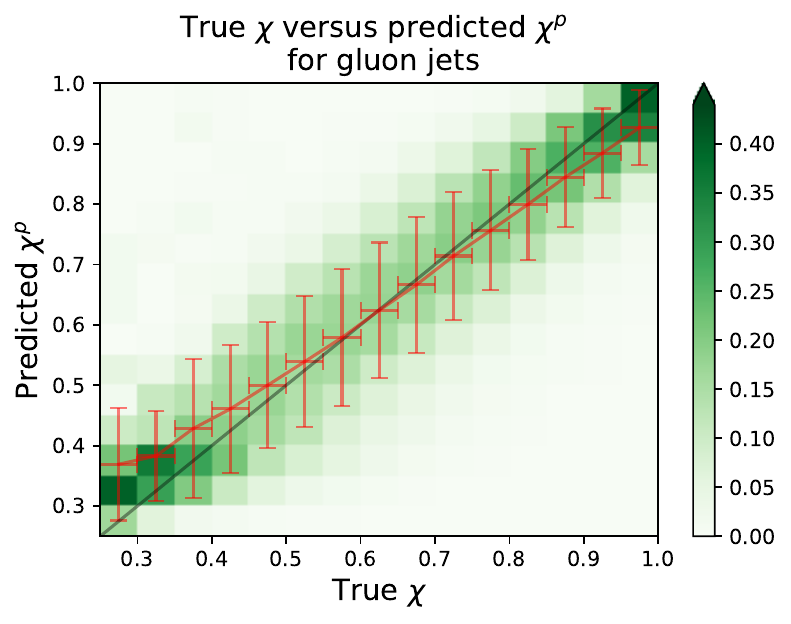}
\caption{CNN prediction performance on quark (left) and gluon (right) jets. The green color represents the probability of predicted $\chi^p$ given true $\chi$ in 2-D histogram, where each column is self-normalized. The red line with error bar quantifies the average and standard deviation of the predicted $\chi^p$ within the given true $\chi$.}
\label{Prediction Performance quark gluon}
\end{figure}
In our previous work~\cite{Du:2020pmp}, we generate approximately 250,000 jets at $\sqrt{s}=5.02$ ATeV for PbPb collisions at 0-5\% centrality within the hybrid strong/weak coupling model~\cite{casalderrey2015erratum}, with the single parameter $\kappa_{\rm sc}$ tuned to reproduce hadron and jet suppression at the LHC \cite{Casalderrey-Solana:2018wrw}. Reconstructed jets with FastJet 3.3.1 
using anti-$k_T$ 
algorithm and $R=0.4$ are required to be within $|\eta|<2$ and to have momentum $p_T^{\mathrm{jet}} > 100\,\, \mathrm{GeV}$. In Fig. \ref{Prediction Performance quark gluon}, we show the prediction performance of energy loss ratio $\chi$ for quark and gluon jets, respectively. One can see that the performance is similar and the network has good compatibility. 



\begin{figure}[h!]
\centering
\includegraphics[width=0.46\textwidth]{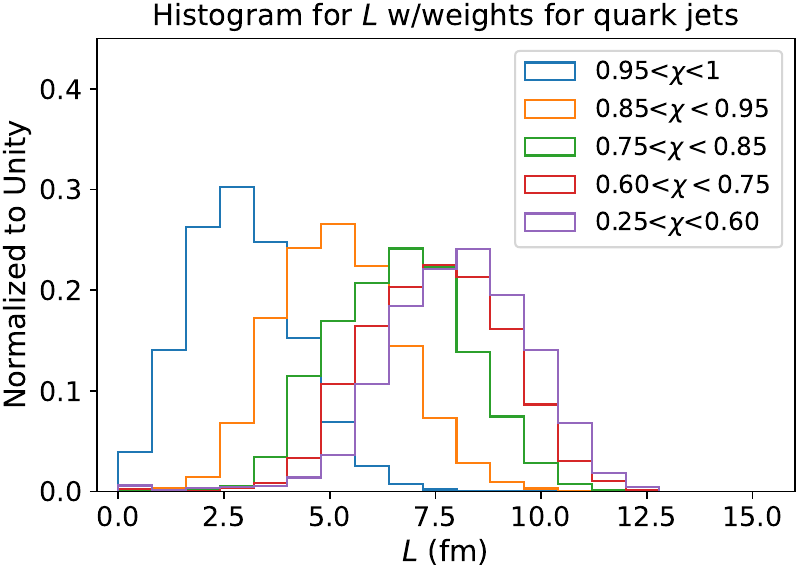}
\includegraphics[width=0.46\textwidth]{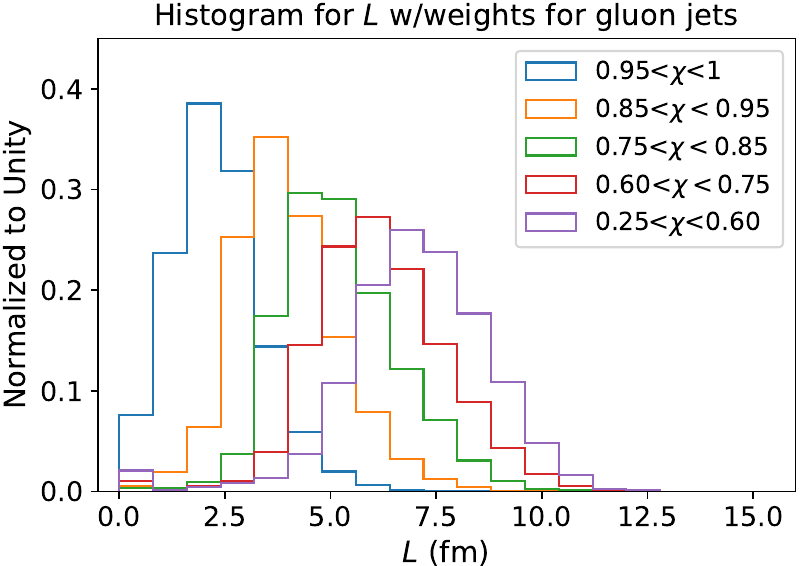}
\caption{Distributions of the in-medium traversed length of quark (left) and gluon (right) jets within different ranges of true values of $\chi$.}
\label{Length}
\end{figure}

The evolution of quark and gluon jets in vacuum is simulated using PYTHIA 8.244.
Their microscopic processes and splitting functions are different. Both the splitting angle as well as the available phase space that can trigger medium-induced radiation are larger, on average, for gluon jets. However, and differently to pQCD computations, the dependence of quark and gluon energy loss in holographic approaches is different only by the ratio of the Casimirs with a reduced power, $(9/4)^{1/3}$~\cite{Gubser:2008as}. Therefore, gluons will lose only slightly more energy when traversing the same length. In Fig.~\ref{Length}, we show the in-medium traversed distance $L$ distribution in the QGP for both quark and gluon jets, sliced in different $\chi$ bins, where $L$ is defined by the traverse momentum $p_{Ti}$ weighted sum of traversed length $L_i$ (in the fluid rest frame) of all the jets constituents, see Ref.~\cite{Du:2020pmp} for details. One can see that jet energy loss increases with the in-medium traversed length. The average energy loss of gluon jets is, as expected, more sensitive to the traversed length. We observe that if we can classify the quark and gluon jets successfully, it will help us pin down the correlation between $L$ and the amount of energy loss, thereby pushing forward our jet tomographic study~\cite{du2021jet}.
Another reason for distinguishing quark/gluon jets is to check the universality of energy loss for different processes, e.g. comparing dijet events (with a mix of quark- and gluon-initiated jets) with boson-jets (where the parton recoiling from the boson is predominantly a quark).



\section{Pre-analysis and Classification Results}
\begin{figure}[tbh]
\centering
\includegraphics[width=0.96\textwidth]{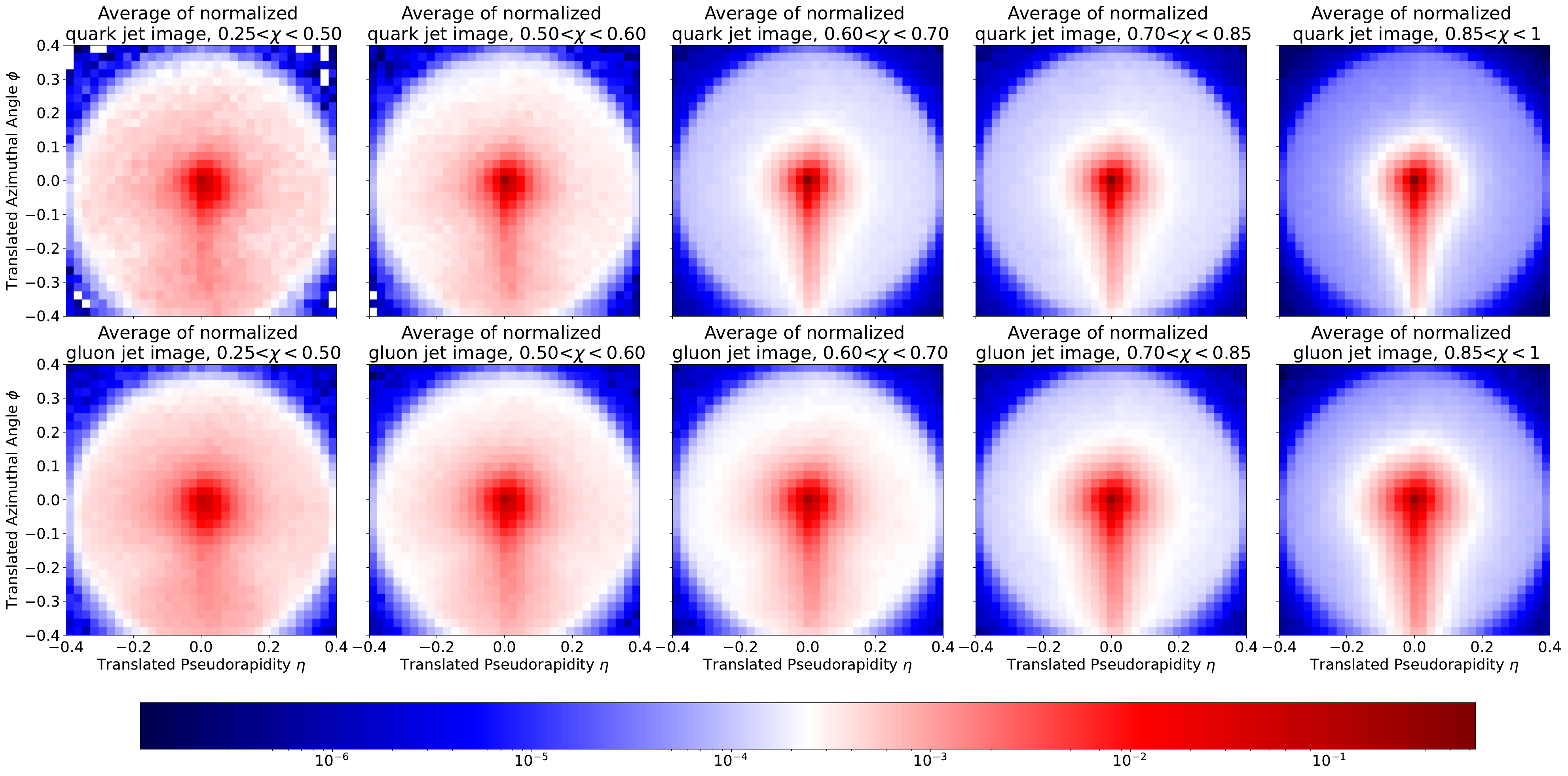}\\
\caption{The average of normalized quark (upper) and gluon (lower) jet image within 5 different $\chi$ cut bins.}
\label{jet image}
\end{figure}
\begin{figure}
\begin{floatrow}
\ffigbox{%
  \includegraphics[width=0.48\textwidth]{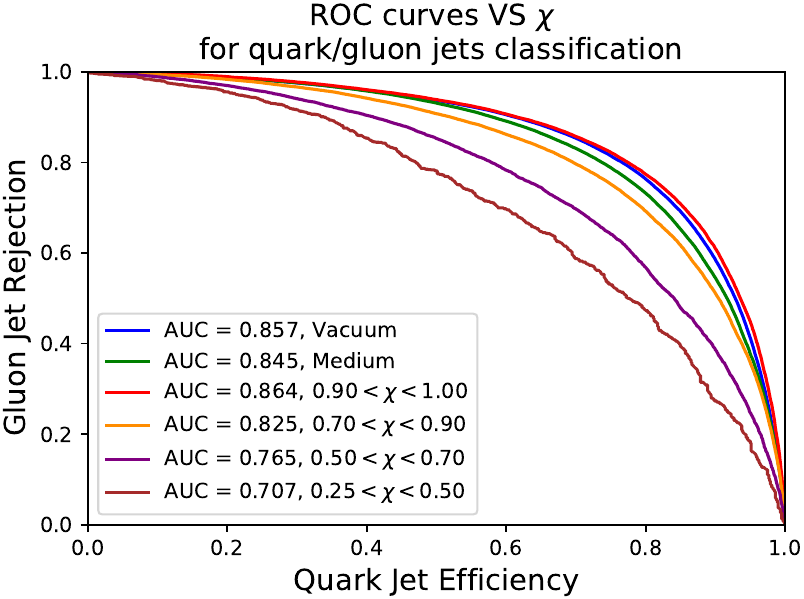}
}{%
  \caption{ROC curves of quark efficiency versus gluon rejection for jets in pp collisions and jets in PbPb collisions for inclusive and sliced in $\chi$ samples.}%
  \label{ROC curves chi}
}
\vspace{-0.3cm}
\linespread{1.5}
\capbtabbox{%
\begin{tabular}{|c|c|c|}
\hline
Input (size)  & Accuracy \\
\hline
Jet shape (8) & 72.2\% \\ 
\hline
JFF (10) & 73.0\%  \\ 
\hline
Jet features (7)&  73.6\% \\ 
\hline
JFF, jet shape (18)   & 74.9\% \\ 
\hline
JFF, jet shape, features (25) & 75.8\% \\ 
\hline
Jet image (33$\times$33) & 75.9\% \\
\hline
\end{tabular}
}{%
  \linespread{1.}
  \caption{Classification performance with different inputs. Jet features include: jet $p_T$, $z_g$, $n_{SD}$, $R_g$, $M$, $M_g$, Multiplicity.}%
  \label{Performance List}
}
\end{floatrow}
\end{figure}

In Fig.~\ref{jet image}, we show the average of $p_T$-normalized quark and gluon jet images in 5 different $\chi$ bins. Here, all the jet images of 33×33 size are rotated and flipped according to certain procedure -- we defer the interested reader to Ref.~\cite{Du:2020pmp}. One can see that with the increase of energy loss, quark and gluon jets show the same qualitative characteristics, i.e., there are more soft particles at large angles within the jet cone, which explains the compatibility of the aforementioned CNN for the energy loss prediction of quark and gluon jets. 
The smearing of the difference of jet substructures poses challenges to the task of classifying quark and gluon jets in the medium.  
In Fig.~\ref{ROC curves chi}, we show the classification performance of quark and gluon jets with CNN, including the results of the vacuum jets as well as the medium jets, inclusive and differential in $\chi$, which are characterized by the receiver operating characteristic curves (ROC curves). The $x$-axis is the quark jet receiving efficiency, while the $y$-axis is the gluon jet rejection efficiency. The identification of quark and gluon jets by the network varies with the classification threshold given by the output of the network. The closer the area under the ROC curves (AUC) is to 1, the more successful the classification task is. Here, the architecture of the CNN is very similar to the one used in Ref.~\cite{Du:2020pmp}.
Overall, the classification accuracy of quark and gluon jets in the medium is a bit lower than that in the vacuum, which agrees with Ref.~\cite{Chien:2018dfn}, and its relative decrease depends on the fraction of the quenched jets. This is consistent with the observation made from the results of the classification for medium jets with different $\chi$, i.e., the greater the energy loss is, the more difficult the classification becomes.

In addition to employing jet images as input, we also use some jet observables and their combinations as inputs to a fully-connected neural network to perform the same classification task for comparison. The architecture of the network is also similar to the one used in Ref. \cite{Du:2020pmp}. These jet observables include jet shape, jet fragmentation function (JFF) and some single-valued jet observables, 
please check Ref.~\cite{Du:2020pmp} for details. The classification performance from these jet observables inputs is listed in Tab.~\ref{Performance List}. One can find that the performance given by the jet shape, jet fragmentation function and jet features increase. This ordering is different from that in the regression task of the energy loss prediction~\cite{Du:2020pmp}, which shows the different sensitivities of these jet observables to different tasks. Similar to the regression task, combining all these observables as input can reproduce the performance given by the jet images, which can serve as an indirect interpretability for the success of the quark/gluon jets classification using the jet images as input.

\section{Summary and Outlook}
In this work, we first showed the good compatibility of the CNN by presenting its prediction performance on the energy loss of quark and gluon jets, respectively, yielding quite similar results. Then we presented the different features of quark and gluon jets after quenching and used deep learning techniques to classify them. It has been found that the greater the energy loss is, the more difficult it is to classify the jets. For now, our study has been purely phenomenological or based on results from selected Monte Carlo models. Given these interesting results, it would also be appealing to pursue a more general theoretical understanding of quark and gluon jet modifications in order to take full advantage of this tagging in the future.




\acknowledgments
This work is supported by the Trond Mohn Foundation under Grant No. BFS2018REK01, the University of Bergen and the Norwegian Research Council grant
255253/F50 - CERN Heavy Ion Theory. Y. D. thanks the support from the Norwegian e-infrastructure UNINETT Sigma2 for the data storage and HPC resources with Project Nos. NS9753K and NN9753K. D. P. has received funding from the European Union’s Horizon 2020 research and innovation program under the Marie Skłodowska-Curie grant agreement No. 754496.

\bibliographystyle{apsrev4-1}
\bibliography{duyl}

\end{document}